\documentclass[twocolumn,showpacs,preprintnumbers,amsmath,amssymb,floatfix]{revtex4}

\RequirePackage{epsfig}
\usepackage{graphicx}

\begin{document}
\title{Two Tunnels to Inflation}
\author{Anthony Aguirre \& Matthew C. Johnson}
\affiliation{
Department of Physics, University of California, Santa Cruz, 
California 95064, USA}

\date{\today}

\begin{abstract}
We investigate the formation via tunneling of inflating (false-vacuum) bubbles in a true-vacuum background, and the reverse process. Using
effective potentials from the junction condition formalism, all true-
and false-vacuum bubble solutions with positive interior and exterior
cosmological constant, and arbitrary mass are catalogued. We find
that tunneling through the same effective potential appears to
describe two distinct processes: one in which the initial
and final states are separated by a wormhole (the Farhi-Guth-Guven
mechanism), and one in which they are either in the same hubble volume or separated by a cosmological horizon. In the zero-mass limit, the first process corresponds to the
creation of an inhomogenous universe from nothing, while the second
mechanism is equivalent to the nucleation of true- or false-vacuum
Coleman-De Luccia bubbles. We compute the probabilities of both mechanisms in the
WKB approximation using semi-classical Hamiltonian methods, and find
that -- assuming both process are allowed -- neither mechanism
dominates in all regimes.
\end{abstract}

\pacs{98.80.Hw, 98.80.Cq}

\maketitle
\section{Introduction}

It has long been appreciated that in a field theory with multiple vacua -- including some models of cosmological inflation -- the nucleation of true-vacuum bubbles in a false-vacuum background can and does occur.  The study of such transitions, with and without gravity, was pioneered by Coleman and collaborators \cite{Coleman:1977py, Callan:1977pt,Coleman:1980aw} and has become a large enterprise.

Real understanding of the {\em reverse} process, nucleation of false-vacuum (inflating) regions in a background of (non-inflating) true-vacuum, has, however, been somewhat more elusive.  It has been proposed that his may occur by the same Coleman-DeLuccia (CDL) instanton responsible for true-vacuum nucleation~\cite{Lee:1987qc,Banks:2002nm,Garriga:1993fh}, by the tunneling of a small false-vacuum bubble through a wormhole to become an inflating region (the Farhi-Guth-Guven, or `FGG' mechanism)~\cite{Farhi:1989yr,Fischler:1990pk,Fischler:1989se}, or by thermal activation~\cite{Garriga:2004nm,Gomberoff:2003zh}. 

This paper comprises the second in a series studying the general process of the nucleation of inflating regions from non-inflating ones. In the first~\cite{Aguirre:2005sv} we cataloged and interpreted all single-bubble thin-wall solutions with an interior false-vacuum de Sitter (`dS') space, and discovered and investigated an instability in such bubbles to non-spherical perturbations. In this paper we attempt to unify the treatment of both false- and true-vacuum bubble nucleations, via the CDL, FGG, and thermal activation mechanisms, in the thin-wall limit.  We find that these can all be studied within 
 a single framework based on the junction condition potentials developed by Guth and collaborators \cite{Farhi:1989yr,Blau:1986cw} and further generalized by Aurilia et. al. \cite{Aurilia:1989sb}~\footnote{The study of thin wall junctions is a vast subject~\cite{Berezin:1982ur,Maeda:1981gw,Ipser:1983db,Aurilia:1984cm,Sato:1986uz,Berezin:1987bc}. In this paper, we will use the notation of Aurilia et. al.~\cite{Aurilia:1989sb}.}. This allows us to both catalog all true- or false-vacuum bubble spacetimes, and to calculate tunneling exponents using the semi-classical Hamiltonian formalism of Fischler et al.~\cite{Fischler:1990pk,Fischler:1989se}.  

Understanding the quantum mechanical~\footnote{It can be shown that this process cannot occur classically unless the weak energy condition is violated~\cite{Farhi:1986ty,Dutta:2005gt,Aguirre:2005sv}.} genesis of inflating regions is very important in assembling a picture of spacetimes containing fields with multiple false vacua, and in understanding how inflation might have begun in our past.  These are related because if inflation can begin from a non-inflating region like our own, then {\em our} inflationary past may have nucleated from non-inflation, and this raises troubling questions~\cite{Dyson:2002pf,Albrecht:2004ke} if spawning inflation is less probable than spawning a large homogeneous big-bang region. This is indeed suggested by singularity theorems showing that inflating false vacuum regions must be larger than the {\em true} vacuum horizon size \cite{Penrose:1964wq,Vachaspati:1998dy} according to some observers \cite{Aguirre:2005sv}. The FGG mechanism provides a potential loophole~\cite{Albrecht:2004ke} because according to an observer in the background true vacuum spacetime, only a region the size of the black hole event horizon is removed.

There have, however, been lingering questions about whether the Farhi-Guth-Guven~\cite{Farhi:1989yr} ``tunneling" process can actually occur. The oldest objection is the fact that the euclidean tunneling spacetime is not a regular manifold~\cite{Farhi:1989yr}. A more modern objection comes from holography: in the FGG mechanism, an observer in the background spacetime only sees a small back hole, whereas the inflating region ``inside" should be described by a huge number of states~\cite{Bousso:2004tv, Banks:2002nm}. This entropy puzzle was recently considered by Freivogel et. al.~\cite{Freivogel:2005qh}, who have used the AdS/CFT correspondence to study thin-walled dS bubbles embedded in a background Schwarzschild-Anti-de Sitter space (Alberghi et. al. have also used the ADS/CFT correspondence to study charged vacuum bubbles~\cite{Alberghi:1999kd}). They find that bubbles containing inflating regions which reside behind a wormhole are represented by mixed states in the boundary field theory. This resolves the entropy puzzle, and also implies that inflating regions hidden behind a wormhole cannot arise from a background spacetime by any unitary process, including tunneling.  It does not, however, suggest why semi-classical methods break down, nor how we should interpret the seemingly-allowed tunneling.

The formalism that we outline in this paper indicates that there are two ways to interpret tunneling through the effective potential of the junction conditions. The existing interpretation (the FGG mechanism) requires that the wall of a false-vacuum bubble (and in some cases of true-vacuum bubbles) must tunnel through a wormhole to produce an inflating region. In this paper, we use the global properties of the Schwarzschild-de Sitter spacetime to show that there is another interpretation corresponding to a mechanism that does not require the existence of a wormhole.

In this mechanism, a small bubble of true- or false-vacuum, which would classically collapse, instead tunnels to a large bubble that exists outside of the {\em cosmological} horizon of the background spacetime. Consequently, this mechanism exists only in spacetimes with a positive cosmological constant. The zero-mass limit of this mechanism correctly reproduces the tunneling exponent for both true- and false-vacuum CDL bubbles~\cite{Coleman:1980aw,Lee:1987qc}. In light of the objections to the FGG mechanism, this new process may be an alternative, in which case the formation of inflating false-vacuum regions by tunneling is forbidden in flat spacetime. On the other hand, these may just be two competing processes, and we will directly compare the tunneling exponents under this assumption.

In section~\ref{classy}, we classify the possible thin-wall true and false one-bubble spacetimes using the effective potential formalism. We then introduce the possible tunneling mechanisms and outline the calculation of the tunneling exponents for the various possibilities in section~\ref{tunneling}. We compare the tunneling rates for the allowed processes in section~\ref{comparison}, interpret our results in section~\ref{bl},  and conclude in section~\ref{conclusions}. 

\section{Classical Dynamics of True and False Vacuum Bubbles}\label{classy}

We will model the true- and false-vacuum bubbles as consisting of a dS interior with cosmological constant $\Lambda_{-}>0$ separated by a thin wall of surface energy density $\sigma$ from a Schwarzschild de Sitter (SdS) exterior with cosmological constant $\Lambda_{+}>0$. If $\Lambda_{-} > \Lambda_{+}$ we will refer to the configuration as a false-vacuum bubble, otherwise it will be denoted a true-vacuum bubble. The exterior metric in the static foliation is given by
\begin{equation}\label{gsds}
ds_{+}^2=-a_{\rm sds}dt^2 + a_{\rm sds}^{-1}dR^2 + R^2 d\Omega^2,
\end{equation}
\begin{equation}\label{defa}
a_{\rm sds}=1-\frac{2M}{R} - \frac{\Lambda_{+}}{3} R^2.
\end{equation}
where $M$ is the usual Schwarzschild mass parameter.
The interior metric in the static foliation is
\begin{equation}\label{gds}
 ds_{-}^2 = -a_{\rm ds} dt^2 + a_{\rm ds}^{-1} dR^2 +R^2 d\Omega^2,
\end{equation}
\begin{equation}
a_{\rm ds}=1- \frac{\Lambda_{-}}{3}R^2,
\end{equation} 

The classical dynamics of thin-walled vacuum bubbles can be determined from the Israel  junction conditions, and the problem has been solved in full generality by Aurilia et. al.  \cite{Aurilia:1989sb}, building on the work of Guth et. al. \cite{Farhi:1989yr,Blau:1986cw}. Assuming spherical symmetry, the radius of curvature of the bubble is the only dynamical variable, so Einstein's equations yield just one equation of motion:
\begin{equation}
\label{israel}
\beta_{\rm ds} - \beta_{\rm sds} = 4 \pi \sigma R,
\end{equation}
where
\begin{equation}\label{betadef}
\beta_{\rm ds} \equiv - a_{\rm ds} \frac{dt}{d\tau},\ \ \beta_{\rm sds} \equiv
a_{\rm sds} \frac{dt}{d\tau}.
\end{equation}
Here, $a$ is the metric coefficient in dS or SdS, and $\tau$ is the proper time of an observer on the bubble wall. The sign of $\beta$
is determined by the trajectory because $dt/d\tau$ could potentially be
positive or negative.

\subsection{Effective potentials}

A set of dimensionless coordinates can be defined, in which
Eq.~\ref{israel} can be written as the equation of motion of a
particle of unit mass in a one dimensional
potential. Let:
\begin{equation} \label{ztor}
z=\left(\frac{L^2}{2M}\right)^{\frac{1}{3}}R,\ \ T = \frac{L^2}{2k} \tau,
\end{equation}
where $M$ is the mass appearing in the SdS metric coefficient, and 
\begin{equation}
k=4\pi\sigma,
\end{equation}
\begin{equation}\label{Lsq}
L^2=\frac{1}{3}\left[\left| \left(\Lambda_{-} + \Lambda_{+} + 3k^2 \right)^2 - 4\Lambda_{+}\Lambda_{-}\right| \right]^{\frac{1}{2}}.
\end{equation}
With these definitions, Eq.~\ref{israel} becomes 
\begin{equation}\label{juncteom}
\left[\frac{dz}{d T }\right]^2=Q-V(z),
\end{equation}
where the potential $V(z)$ and energy $Q$ are 
\begin{equation}\label{potential}
V(z)=-\left[z^2+\frac{2Y}{z}+\frac{1}{z^4} \right],
\end{equation}
with
\begin{equation} \label{y}
Y=\frac{1}{3}\frac{\Lambda_{+}-\Lambda_{-}+3k^2}{L^2},
\end{equation}
and
\begin{equation} \label{Qtom}
Q=-\frac{4k^2}{\left(2M\right)^{\frac{2}{3}}L^{\frac{8}{3}}}.
\end{equation}
Note that a small negative $Q$ corresponds to a large mass, so that
even between $-1 < Q < 0$ the mass can be arbitrarily large.
The scale of all quantities of interest is set by some power of the bubble wall surface energy density ($k$) if the interior and exterior cosmological constants are written in terms of $k^2$ as
\begin{equation}
\Lambda_{+}=Ak^2, \ \ \Lambda_{-}=Bk^2.
\end{equation}

\begin{figure*}[h!]
\centering
\includegraphics[width=17cm]{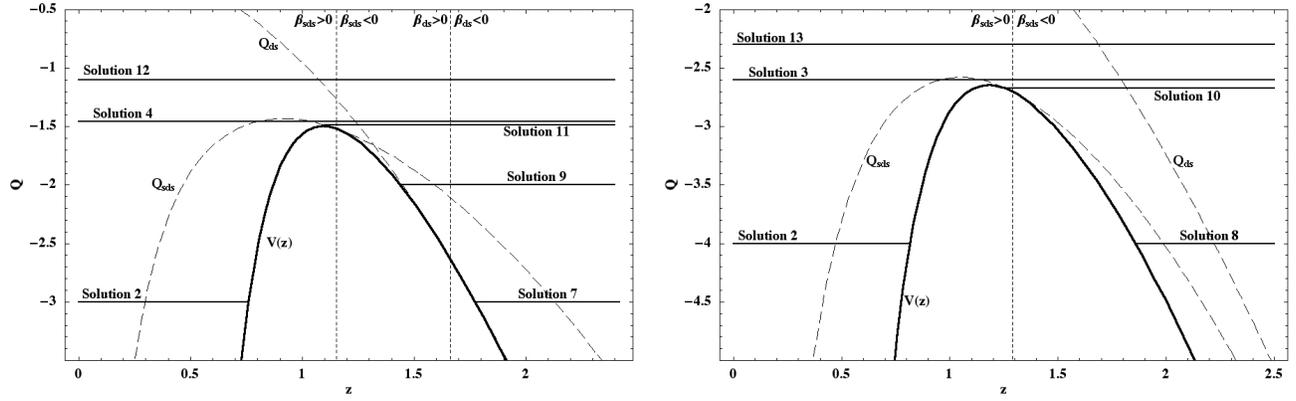}
\caption{Potential for false-vacuum bubbles with $B < 3(A-1)$. The diagram on the left is for ($A=9$, $B=15$). The diagram on the right is for ($A=2.9$, $B=3$), which is an example of a case where there is no $\beta_{\rm ds}$ sign change ($ B < A + 3 < 3 (A - 1) $).  The two dashed lines labeled $Q_{\rm sds}$ and $Q_{\rm ds}$ represent the exterior and interior horizon crossings respectively. The vertical dotted lines denote the regions in which $\beta_{\rm sds}$ and $ \beta_{\rm ds}$ are positive and
negative. Various trajectories are noted. \label{Bgt3Am1}}
\end{figure*}

From the constant-$Q$ trajectories in the presence of the potential of Eq.~\ref{potential}, one can construct the full one-bubble spacetimes \cite{Blau:1986cw,Farhi:1989yr, Aguirre:2005sv}. Shown in Fig.~\ref{Bgt3Am1} is an example of two of the possible potential diagrams. In addition to the potential Eq.~\ref{potential}, there are other landmarks in Fig.~\ref{Bgt3Am1}. Intersections with the dashed line $Q_{\rm ds}$ (which is obtained by solving $a_{\rm ds}=0$ for $Q$) as one moves along a line of constant $Q$ represent a crossing of either the past or future horizon of the interior dS spacetime. Every intersection with the dashed line $Q_{\rm sds}$ represents a horizon crossing in the SdS spacetime (this could represent either the past- or future-black hole {\em or} cosmological horizons). It can be shown~\cite{Aurilia:1989sb} that $\beta_{\rm ds}$ and $\beta_{\rm sds}$ are monotonic functions of $z$, which will have zeros where $Q_{\rm ds}$ or $Q_{\rm sds}$ intersect the potential. These points demarcate sign changes in $\beta_{\rm ds}$ or $\beta_{\rm sds}$, and are denoted by the vertical dotted lines in Fig.~\ref{Bgt3Am1}.   

\begin{figure*}
\centering
\includegraphics[width=17cm]{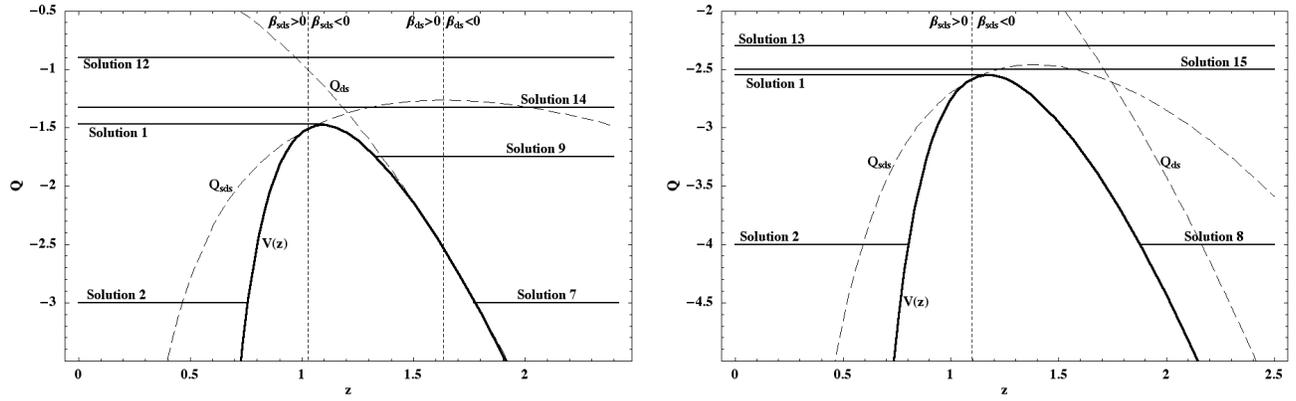}
\caption{Potential for false-vacuum bubbles with $B > 3(A-1)$. The diagram on the left is for ($A=1$, $B=6$). The diagram on the right is for ($A=1$, $B=2$), which is an example of a case where there is no $\beta_{\rm ds}$ sign change ($3 (A-1) < B < A + 3$). For these choices of parameters, the sign change in $\beta_{\rm sds}$ occurs to the left of the maximum in the potential. Various trajectories are noted. \label{Blt3Am1}}
\end{figure*}

\begin{figure*}
\centering
\includegraphics[width=17cm]{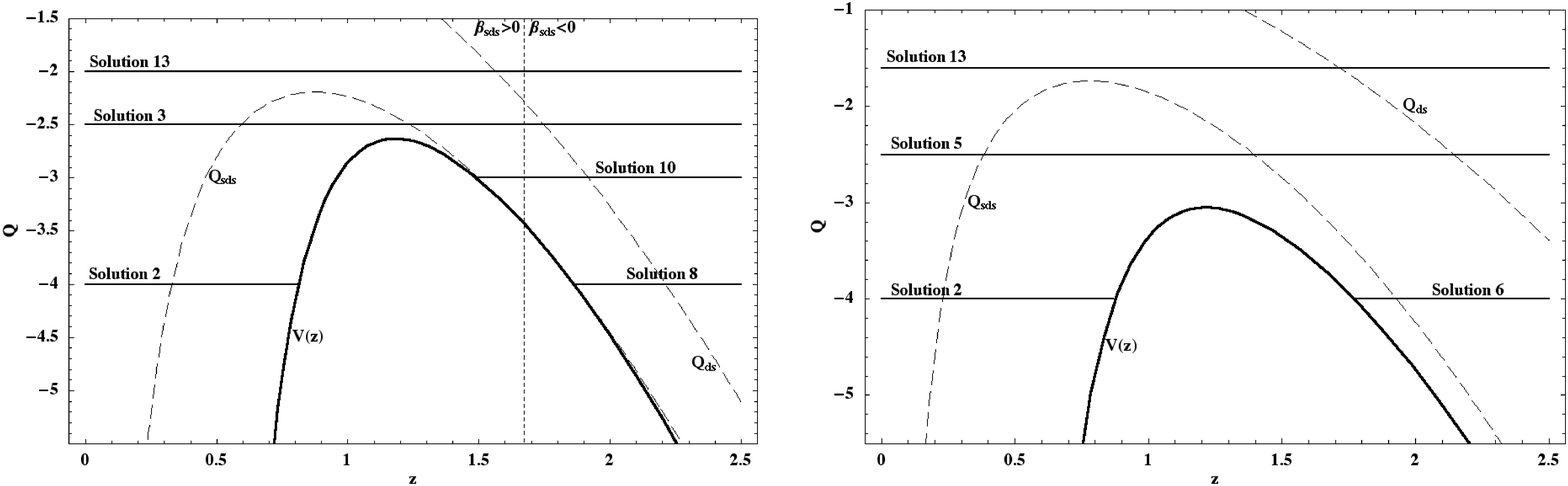}
\caption{Potential for true-vacuum bubbles with $A > \frac{B}{3} + 1$. The diagram on the left is for ($A=7$, $B=6$), which is an example of a case where there is a $\beta_{\rm sds}$ sign change ($A < B+3$). The diagram on the right is for ($A=14$, $B=8$), which contains no $\beta_{sds}$ sign change ($A > B+3$).  Various trajectories are noted. 
\label{AgtBo3p1}}
\end{figure*}

\begin{figure}
\begin{center}
\epsfig{file=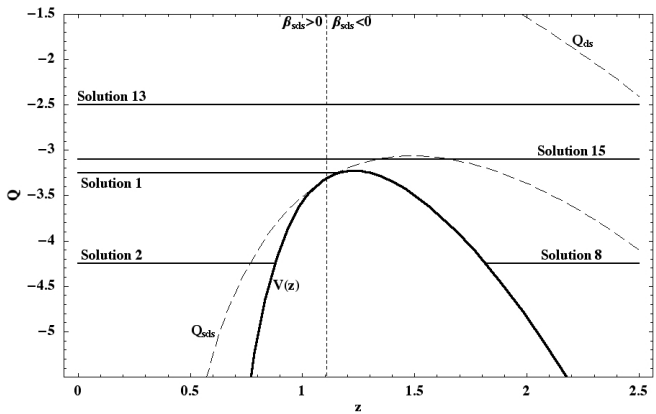,width=8.6cm}
\caption{Potential for true-vacuum bubbles with ($A=.6$, $B=.5$), corresponding to the case where $A < \frac{B}{3}+1 < B+3$. Various trajectories are noted. \label{A_6B_5}}
\end{center}
\end{figure}

For there to be a $\beta_{\rm ds}$ sign change, $Y$ in Eq.~\ref{y} must be in the range $-1 \leq Y < 0$ \cite{Aurilia:1989sb}, which yields the condition that $B > A+3$ if a sign change is to occur. This inequality shows that $\beta_{\rm ds}$ does not change sign for true vacuum bubbles ($A>B$). For there to be a $\beta_{\rm sds}$ sign change, the function
\begin{equation}
\tilde{Y} = \frac{1}{3}\frac{\Lambda_{+}-\Lambda_{-}-3k^2}{L^2}
\end{equation}
must be in the range $-1 \leq \tilde Y < 0$ \cite{Aurilia:1989sb}, which yields the condition that $B > A-3$ if a $\beta_{\rm sds}$ sign change is to occur. If a $\beta_{\rm sds}$ sign change does exist, it can occur to the left (if $B > 3(A-1)$)
or right ($B < 3 (A-1)$) of the maximum in the potential \cite{Aguirre:2005sv}. Given these conditions, there are a total of seven qualitatively different potential diagrams to consider, examples of which are shown in Figs.~\ref{Bgt3Am1}, \ref{Blt3Am1}, \ref{AgtBo3p1}, and \ref{A_6B_5}. 

\subsection{Conformal diagrams}

\begin{figure*}
\begin{center}
\includegraphics[width=14cm]{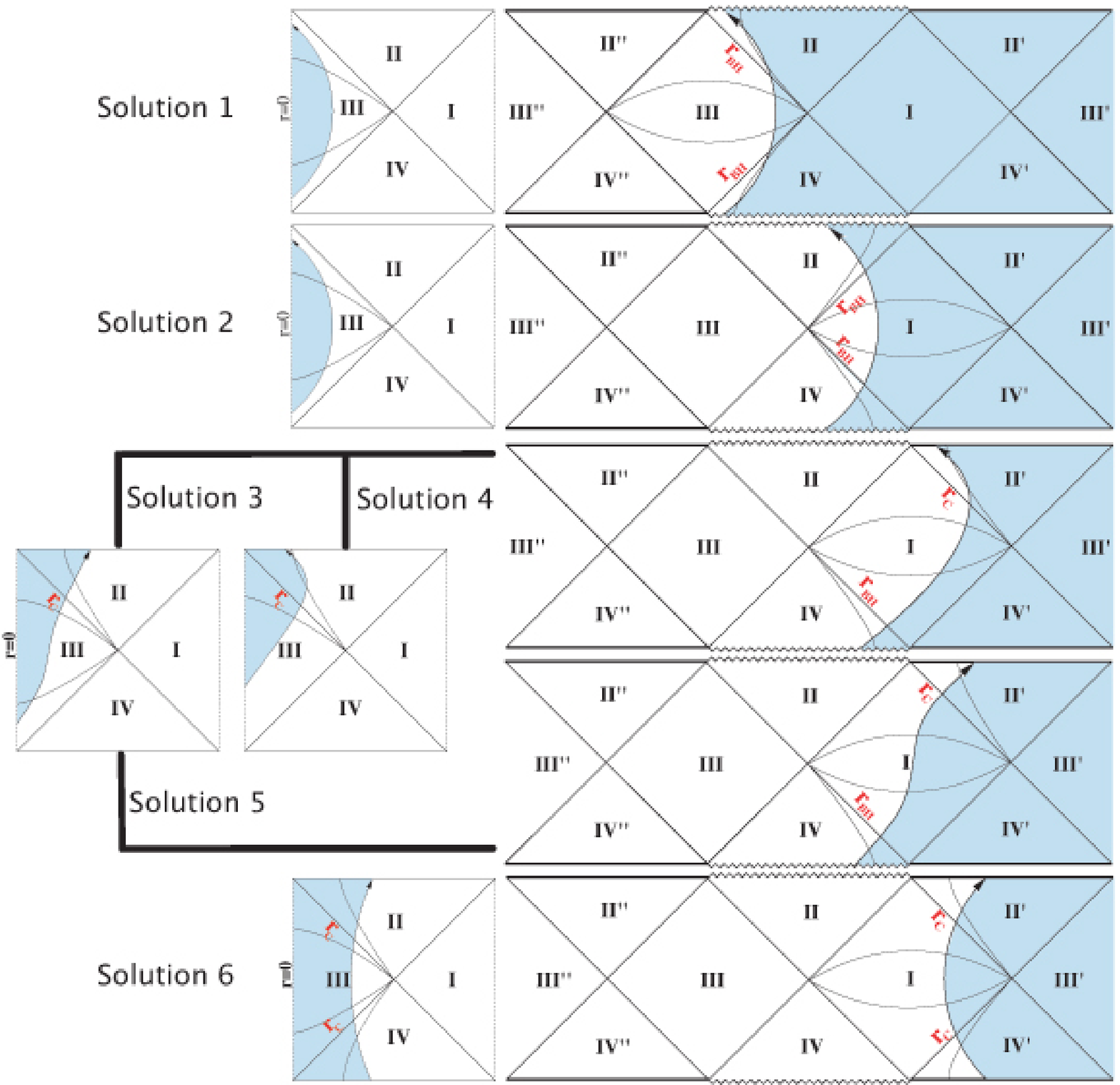}
\caption{Conformal diagrams for the one-bubble spacetimes which do not lie behind a worm hole. The global one-bubble spacetimes are constructed by matching the interior (shaded regions of the dS conformal diagrams in the left column) to the exterior (shaded regions of the SdS conformal diagrams in the right column) across the bubble wall (solid line with an arrow). For solutions with qualitatively similar SdS diagrams, the various options for the dS interior are shown.
\label{diags1}}
\end{center}
\end{figure*}

\begin{figure*}[htbp]
\begin{center}
\includegraphics[width=16cm]{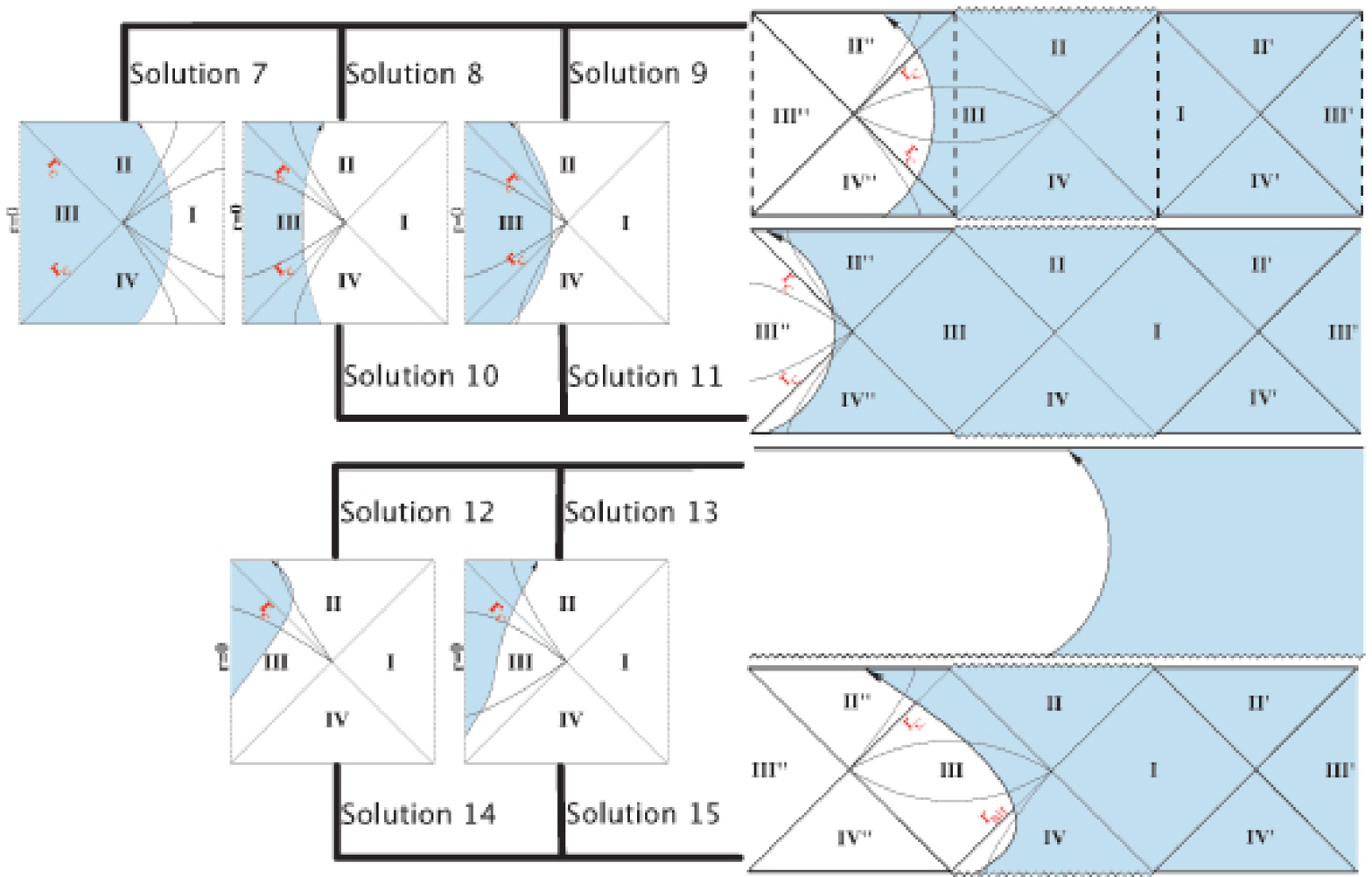}
\caption{Conformal diagrams for the one-bubble spacetimes which lie behind a worm hole. The global one-bubble spacetimes are constructed by matching the interior (shaded regions of the dS conformal diagrams in the left column) to the exterior (shaded regions of the SdS conformal diagrams in the right column) across the bubble wall (solid line with an arrow). For solutions with qualitatively similar SdS diagrams, the various options for the dS interior are shown.
\label{diags2}}
\label{default}
\end{center}
\end{figure*}

\begin{figure*}
\begin{center}
\includegraphics[width=12cm]{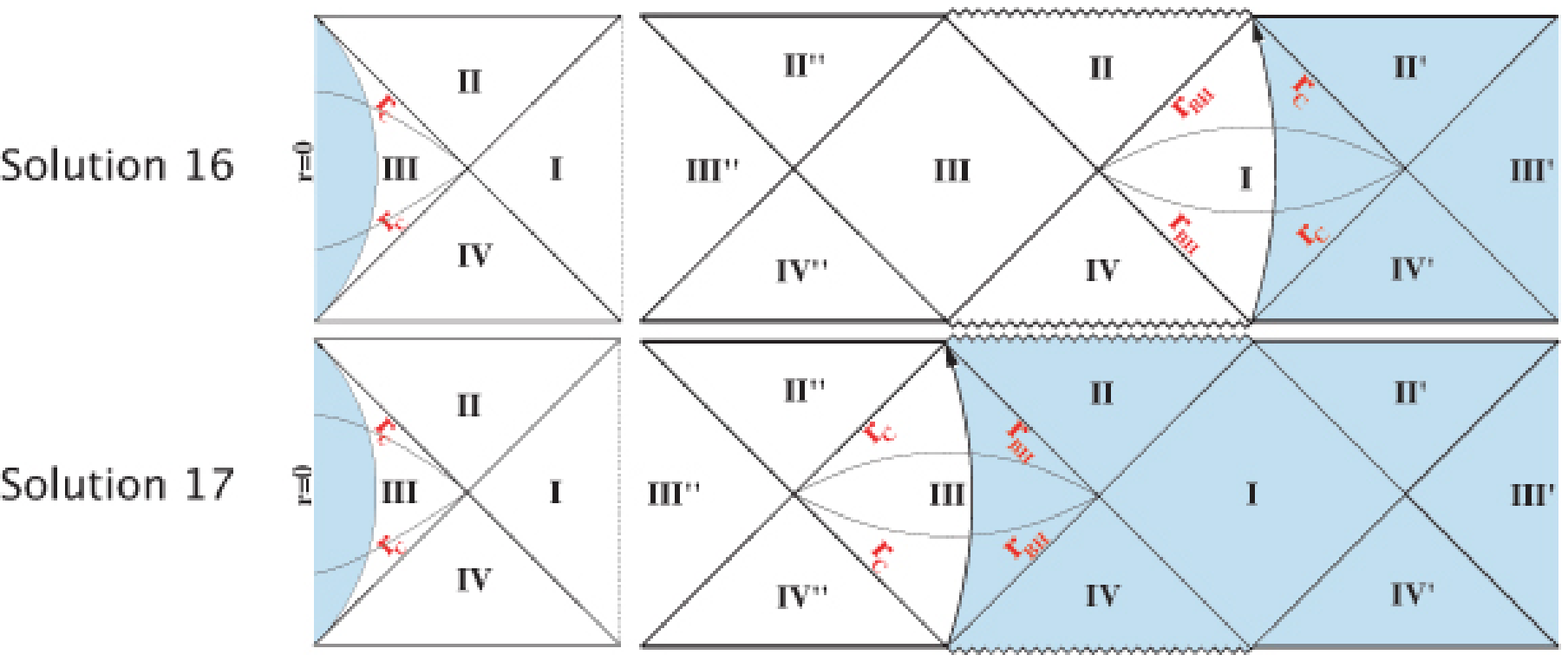}
\caption{Solutions which are in unstable equilibrium between the bound and unbound solutions of Fig.~\ref{diags1} and \ref{diags2}. These solutions correspond to the time symmetric spacetimes of thermally activated bubbles. \label{thermalons}}
\label{default}
\end{center}
\end{figure*}

The one-bubble spacetimes, represented by lines of constant $Q$ on the junction condition potential diagrams, are shown in Figs.~\ref{diags1}, \ref{diags2}, and \ref{thermalons}~\footnote{Many of these solutions have appeared in previous work~\cite{Blau:1986cw,Sato:1986uz,Berezin:1987bc,Aurilia:1989sb,Gomberoff:2003zh,Garriga:2004nm,Aguirre:2005sv}, but with specific assumptions about the mass and/or the interior and exterior cosmological constants.}. The shaded regions of the conformal diagrams shown in the left column cover the interior of the vacuum bubble. The shaded regions of the diagrams in the right column cover the spacetime outside the bubble. The conformal diagrams in each row are matched along the bubble wall (solid line with an arrow). For solutions with qualitatively similar SdS diagrams, the various options for the dS interior are connected by labeled solid lines.

The conformal diagrams shown in Fig.~\ref{diags1} are all solutions in which the bubble wall remains to the right of the wormhole of the SdS conformal diagram. The bound solutions, Solutions 1 and 2, exist for both true- and false-vacuum bubbles. For false-vacuum bubbles, they represent a regime in which the inward pressure gradient and bubble wall tension dominate the dynamics, causing the bubble to ultimately contract. In the case of true-vacuum bubbles, this corresponds to cases where the wall tension overwhelms the outward pressure gradient. 

In the monotonic Solutions 3-5 of Fig.~\ref{diags1} the bubble wall has enough kinetic energy to reach curvatures comparable to the exterior horizon size, at which time the bubble cannot collapse. Solutions 3 and 4 represent either true- or false-vacuum bubbles where the wall tension and/or the inward pressure gradient causes the wall to accelerate towards $r=0$, but which are saved from collapse by the expansion of the exterior spacetime. Solution 5 exists only for true-vacuum bubbles, and describes a solution which accelerates away from the origin due to the outward pressure gradient while also being pulled out of the cosmological horizon by the expansion of the exterior spacetime. 

The unbound Solution 6 also exists only for true-vacuum bubbles. Here, the bubble expands, all the while accelerating towards the false-vacuum. The zero mass limit ($M \rightarrow 0$, or $Q \rightarrow -\infty$) of this solution is the one-bubble spacetime of the analytically continued true-vacuum Coleman-De Luccia (CDL) instanton \cite{Coleman:1980aw} in the limit of an infinitely thin wall. This can be seen by considering the limit as the potential (Eq.~\ref{potential}) goes to $-\infty$, where on the right (unbound) side of the potential hump the $z^2$ term dominates. Solving for $R$ using Eq.~\ref{ztor}, we find the radius at turnaround to be
\begin{equation}\label{r_0inst}
R = 6k \left[\left|\left(\Lambda_{+} + \Lambda_{-} +3k^2 \right)^2 - 4\Lambda_{+}\Lambda_{-} \right| \right]^{-1/2},
\end{equation} 
which is indeed the radius of curvature of the CDL instanton \cite{Coleman:1980aw} at nucleation. 

The solutions shown in Fig.~\ref{diags2} are all behind the wormhole in the SdS spacetime, save Solutions 12 and 13, which correspond to evolution in a spacetime without horizons. The false-vacuum bubble solutions 7 and 9, and true- or false-vacuum bubble solution 8 are unbound solutions which exist to the left of the worm hole on the SdS conformal diagram.  It can be seen that at turnaround, each of these bubbles will be larger than the exterior horizon size. Observers in region III of the SdS conformal diagram will see themselves sandwiched between a black hole and a bubble wall which encroaches in from the cosmological horizon. Observers inside the bubble are also surrounded by a bubble wall, and so we are faced with the rather odd situation that both observers will perceive themselves inside bubbles of opposite phase. 

Solutions 7 and 8 have interesting zero mass limits. Since these solutions involve both sides of the wormhole, the zero mass limit corresponds to an exactly dS universe consisting of regions I, II', III', and IV' (encompassed by the vertical dashed lines shown on the right side of the first diagram of Fig.~\ref{diags2}) of the SdS diagram (in which nothing happens), and a dS universe consisting of regions III, II'', and IV'' (encompassed by the other set of vertical dashed lines) which contains a CDL true- or false-vacuum bubble. The radius at the turning point is still given by Eq.~\ref{r_0inst}, and so the bubble to the left of the wormhole is the analytic continuation of the true- or false-vacuum CDL instanton. However, note that the Lorentzian evolution of the true-vacuum bubbles is very different from the canonical CDL instanton discussed in the previous paragraph. As seen from the outside (region III of the SdS diagram on the right), the bubble wall accelerates towards the true-vacuum (driven by the wall tension); in the absence of the cosmic expansion of the false-vacuum, this solution would be bound.

\begin{figure}
\begin{center}
\includegraphics[width=8.6cm]{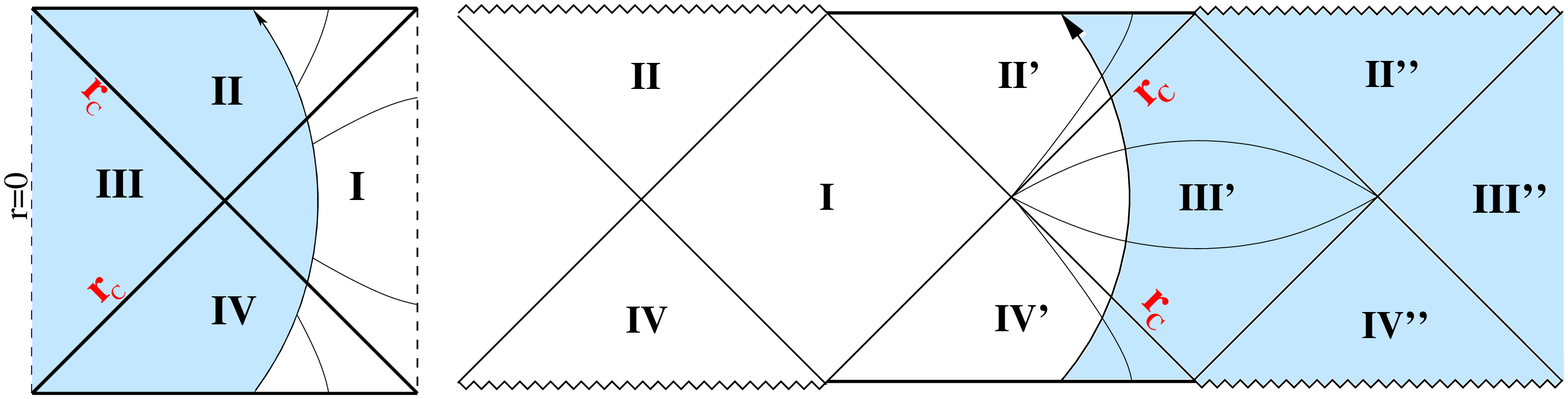}
\caption{Solutions can be to the right of region I instead of behind the wormhole. This solution is identical to Solution 7 of Fig.~\ref{diags2}. \label{noworm}}
\label{default}
\end{center}
\end{figure}

Because the SdS manifold is non-compact, there are actually many more options. We have so far placed special significance on the singularities in regions II and IV of the SdS diagram. However, there will be other singularities both to the left and right of these regions which can also be viewed as the origin of coordinates. It is perfectly legitimate to construct bubble wall solutions using any origin of coordinates one wishes, and therefore each of the solutions in Fig.~\ref{diags1} and \ref{diags2} represents only one of an infinity of possible solutions. An example of an alternative solution is shown in Fig.~\ref{noworm}, which is identical to the Solution 7 in Fig.~\ref{diags2} in every way, except different regions of the conformal diagram are physical. This observation is key for the tunneling mechanisms we will describe in the next section.

Moving on to the other solutions in Fig.~\ref{diags2}, Solution 10 (corresponding to either true- or false-vacuum bubble) and Solution 11 (corresponding to a false-vacuum bubble) are massive unbound solutions which lie outside the cosmological horizon of a region III observer. Solution 12 (corresponding to a false-vacuum bubble) and Solution 13 (corresponding to either a true- or false-vacuum bubble) are monotonic solutions with mass greater than the Nariai mass of the SdS spacetime. This can be seen by noting that these constant $Q$ trajectories never cross the $Q_{\rm sds}$ line in the potential diagrams. The false-vacuum bubble Solution 14, and the true- or false-vacuum bubble solution 15 are monotonic solutions which must lie to the left of the wormhole. 

There is one more class of solutions, shown in Fig.~\ref{thermalons}, which exist in unstable equilibrium between the bound and unbound solutions of Fig.~\ref{diags1} and \ref{diags2}. Solution 16 corresponds to true- or false-vacuum bubbles with $B < 3 (A-1)$, while Solution 17 corresponds to true- or false-vacuum bubbles with $B > 3 (A-1)$. These solutions can be identified as the spacetimes of the thermal activation mechanism of Garriga and Megevand~\cite{Garriga:2004nm}, which we will discuss further in Sec.~\ref{highlow} and~\ref{comparison}.

Classical trajectories exist on either side of the potential diagrams of Figs.~\ref{Bgt3Am1}, \ref{Blt3Am1}, \ref{AgtBo3p1}, and \ref{A_6B_5}, and so one can ask if there is any quantum process that connects two solutions of the same mass through the classically forbidden region under the potential. This would correspond to transitions from the bound spacetimes shown in Fig.~\ref{diags1} (Solutions 1 and 2) to the unbound spacetimes shown in Figs.~\ref{diags1} and \ref{diags2} (Solutions 6-11). Such processes do seem to occur \cite{Farhi:1989yr,Fischler:1989se,Fischler:1990pk,Berezin:1988dz,Khlebnikov:2003ve}, at least within the framework of semi-classical quantum gravity, and we now turn to the problem of determining which transitions are allowed and with what probabilities.

\section{Tunneling}\label{tunneling}

The potential diagrams discussed in the previous section nicely summarize the classically allowed one-bubble spacetimes. They also illustrate the possibility that there might exist some process akin to the tunneling of a point particle through a potential barrier. Such a process would correspond to the quantum tunneling between thin-wall bubbles of equal mass, but different turning-point radii. We will find that in SdS, there are actually two different semi-classical tunneling processes which connect equal mass solutions: the FGG mechanism \cite{Farhi:1989yr} and a process which is only allowed in the presence of an exterior cosmological constant, and which has a zero-mass limit that corresponds to CDL true- or false-vacuum bubble nucleation~\cite{Coleman:1980aw, Lee:1987qc}. 

\subsection{Hamiltonian formalism}

In a pair of papers, Fischler et. al. (FMP) \cite{Fischler:1990pk,Fischler:1989se} presented a calculation of the probability for transitions between various thin-wall false-vacuum bubble solutions. This calculation was done using Hamiltonian methods in the WKB approximation for the case where the exterior cosmological constant is zero. A similar calculation of such tunneling events was performed by Farhi et. al. \cite{Farhi:1989yr} using a path integral approach. Both methods encounter the difficulty that the interpolating geometry involves a two-to-one mapping to the exterior spacetime, and thus  is {\em not} a manifold. We will use the Hamiltonian approach, which is the most direct route to a tunneling exponent and temporarily skirts this issue. A discussion of the interpolating geometry will appear in a forthcoming publication~\cite{Aguirre:xi}.

Here, we extend the calculation of FMP to include all spacetimes with arbitrary non-negative interior ($\Lambda_{-}$) and exterior ($\Lambda_{+}$) cosmological constants. This formalism, with the catalog of all classically allowed solutions, will allow us to create a complete listing of the possible tunneling events. 

Following FMP, we begin by making a coordinate transformation to recast the interior and exterior metrics in Eqs.~\ref{gsds} and \ref{gds} into the form
\begin{eqnarray}\label{metric}
ds^2 &=& - N^{t}\left(t,r\right)^2 dt^2 + L\left(t,r\right)^2 \left[dr + N^{r}\left(t,r\right) dt \right]^2 \nonumber \\ && + R\left(t,r\right)^2 \left(d\theta^2 + \sin^2 \theta d\phi^2 \right),
\end{eqnarray}
where $N^{t}\left(t,r\right)$ is the lapse function, $N^{r}\left(t,r\right)$ is the shift, and $L \equiv ds/dr$. The action for a general theory of matter coupled to gravity is then given by
\begin{equation}
S = \int dt \ p \ \dot{q} + \int dr \ dt  \ \left(\pi_{L} \dot{L} + \pi_{R} \dot{R} - N^{t} H_{t} - N^{r} H_{r} \right)
\end{equation}
where $\pi_{L}$ is the momentum conjugate to $L$, and $\pi_{R}$ is the momentum conjugate to $R$. This action, with the four constraints
\begin{subequations}\label{constraints}
\begin{equation}
H_{t, r} \left(q, L, R, p, \pi_{L}, \pi_{R} \right)=0,
\end{equation}
\begin{equation}
\pi_{N^{t}}=\pi_{N^{r}}=0,
\end{equation}
\end{subequations}
fully determines the classical evolution of the system. For a thin-walled bubble with an arbitrary surface energy density $k$ and interior and exterior cosmological constant ($\Lambda_{-}$ and $\Lambda_{+}$), the Hamiltonian densities are given by
\begin{eqnarray}\label{Ht}
H_{t} &=& \frac{L \pi_{L}^{2}}{2 R^2} - \frac{\pi_{L} \pi_{R}}{R} \nonumber \\  &&+ \frac{1}{2} \left[ \left[\frac{2RR'}{L}\right]' - \frac{R'^2}{L} - L + \Lambda_{+} L R^2 \right] \nonumber \\ &&+ \Theta\left(r_{w} -r \right) \frac{\left(\Lambda_{-} - \Lambda_{+} \right)}{2} L R^2  \nonumber \\ && + \delta\left(r_{w}-r\right) \left(L^{-2} p_{w}^{2} + k^2 R_{w}^{4}  \right)^{1/2},
\end{eqnarray}
\begin{equation}\label{Hr}
H_{r} = R' \pi_{R} - L \pi_{L}' - \delta\left(r_{w} - r \right) p_{w},
\end{equation}
where a prime denotes a derivative with respect to $r$ and $r_{w}$ is the position of the bubble wall (quantities with the subscript $w$ are evaluated at this position). 

A linear combination of the constraints Eq.~\ref{Ht} and \ref{Hr} can be use to eliminate $\pi_{R}$
\begin{equation}
\frac{R'}{L} H_{t} + \frac{\pi_{L}}{RL} H_{r} = 0,
\end{equation}
which, if we define 
\begin{equation}\label{M}
\mathcal{M} \equiv  \frac{\pi_{L}^{2}}{2R} + \frac{R}{2} \left[1 - \left[\frac{R'}{L}\right]^2 - \frac{\Lambda_{\pm} R^2}{3}   \right],
\end{equation}
can be written as
\begin{equation}\label{M'}
\mathcal{M}' = \delta\left(r_{w}-r\right) \left(\frac{R'}{L} \left(L^{-2} p_{w}^{2} + k^2 R_{w}^{4}  \right)^{1/2} + \frac{\pi_{L}}{RL} p_{w} \right).
\end{equation}
It can be seen from Eq.~\ref{M'} that $\mathcal{M}$ is zero for $r < r_{w}$ and independent of $r$ for $r > r_{w}$. 
We will define $\mathcal{M}(r>r_{w}) \equiv M$, which is the mass enclosed by a surface with $r>r_{w}$. Solving for $\pi_{L}$ at $r=0$ and $r=\infty$ using the conditions on $\mathcal{M}$ yields:
\begin{equation}\label{piLsmall}
\pi_{L}^{2} = - R^2 \left[ 1 - \left[\frac{R'}{L} \right]^2 - \frac{\Lambda_{-}R^2}{3}  \right], \ \ \ \ \ r < r_{w}
\end{equation}
\begin{equation}\label{piLbig}
\pi_{L}^{2} = - R^2 \left[ 1 - \left[\frac{R'}{L} \right]^2 - \frac{\Lambda_{+}R^2}{3}  - \frac{2M}{R} \right], \ \ \ \ \ r > r_{w}.
\end{equation}
From $H_{r} = 0$, solving for $\pi_{L}'$, and integrating from $r_{w} - \epsilon$ to $r_{w} + \epsilon$, one finds that the discontinuity in $\pi_{L}$ across the wall ($\Delta \pi_{L}$) is
\begin{equation}
\Delta \pi_{L} = - \frac{p_{w}}{L_{w}},
\end{equation}
From $H_{t} = 0$, solving for $R''$, and integrating from $r_{w} - \epsilon$ to $r_{w} + \epsilon$, one finds that the discontinuity in $R'$ across the wall ($\Delta R'$) is
\begin{equation}
\Delta R' = - \frac{1}{R_{w}} \left[ p_{w}^2 + k^2 L^2 R_{w}^4 \right].
\end{equation}
These discontinuity equations are equivalent to the Israel junction conditions, and can be manipulated to reproduce Eq.~\ref{israel}. There are classically allowed and forbidden regions in the space of $R$, $L$, and $r$, the boundaries between which can be found by looking for where the conjugate momenta are zero. There is, however, only one true degree of freedom, the classically allowed/forbidden region for which is classified by the potential Eq.~\ref{potential}. The unphysical degrees of freedom will allow for a variety of physically equivalent paths through the  the space of $(L, R, r)$.

To quantize the system, we impose the constraints of Eq.~\ref{constraints} on the wave functional $\Psi$:
\begin{equation}
\hat{H}_{t} \Psi=\hat{H}_{r} \Psi=\hat{\pi}_{N^{t}}\Psi=\hat{\pi}_{N^{r}}\Psi=0.
\end{equation}
The last two constraints restrict the wave functional to depend only upon $L$, $R$, and $r$, which in the WKB approximation is taken to be
\begin{equation}\label{psi}
\Psi\left(L,R,r\right) = \exp \left[ i \Sigma_{0} \left(L,R,r \right) / \hbar + O\left(\hbar \right)\right].
\end{equation}
We explicitly include $\hbar$ here to emphasize the order of our approximation, but note that we use geometrical units in all other cases. Acting with $\hat{H}_{t}$ and $\hat{H}_{r}$, and keeping terms in the Taylor expansion only to leading order in $\hbar$ (which removes any operator ordering ambiguities) yields the Hamilton-Jacobi equations 
\begin{equation}
H_{r,t} \left(r,L,R,\frac{\delta  \Sigma_{0}}{\delta r}, \frac{\delta  \Sigma_{0}}{\delta L},\frac{\delta  \Sigma_{0}}{\delta R}  \right) = 0.
\end{equation}
We will integrate
\begin{equation}\label{dSigma}
\delta \Sigma_{0} = \hat{p} \delta \hat{r} + \int_{0}^{\infty} dr \left[ \pi_{L} \delta L + \pi_{R} \delta R    \right],
\end{equation}
to solve for the exponent of the wave functional Eq.~\ref{psi}.

\subsection{Calculating tunneling rates}

\begin{figure}
\begin{center}
\epsfig{file=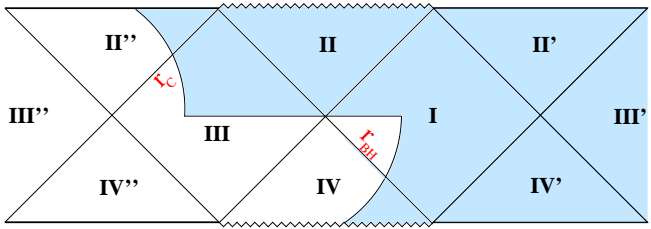,width=8.6cm}
\caption{The FGG mechanism: tunneling from a bound solution to an unbound solution on the other side of a wormhole. \label{tunnel}}
\end{center}
\end{figure}

The problem that we wish to solve is the tunneling amplitude in the WKB approximation to connect bound solutions with turning point $R_{1}$ to equal-mass unbound solutions with turning point $R_{2}$. An example of this is the FGG mechanism \cite{Farhi:1989yr}, which consists of two steps. First, an expanding region of false/true-vacuum, which would classically collapse into a black hole, is formed and evolves to the classical turning point. Here, there is a chance for the bubble wall to tunnel through the wormhole to one of the unbound solutions, as shown in Fig.~\ref{tunnel}. The result of this process is a black hole in the region of the old phase, which is connected by a wormhole to a universe containing an expanding bubble of the new phase.

\begin{figure}
\begin{center}
\epsfig{file=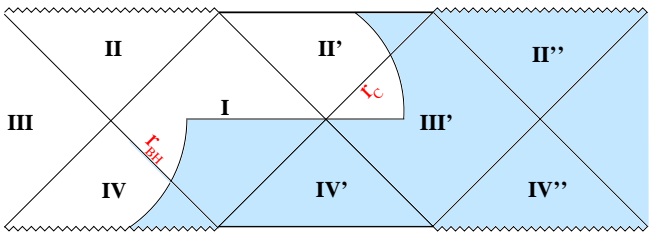,width=8.6cm}
\caption{Tunneling from a bound solution to an unbound solution which exists outside the cosmological horizon. \label{tunnelnoworm}}
\end{center}
\end{figure}

As we saw in Sec.~\ref{classy}, because SdS is non-compact, there are many possible one-bubble spacetimes where region I of the SdS conformal diagram is not physical. We can therefore imagine tunneling from the bound Solution 1 or Solution 2 of Fig.~\ref{diags1} to the unbound spacetime shown in Fig.~\ref{noworm}. This process, which can occur only in the presence of a a positive exterior cosmological constant, is depicted in Fig.~\ref{tunnelnoworm}. For every transition which goes through the wormhole, as in the FGG mechanism, there is another transition which instead goes out the cosmological horizon.

There are many possible transitions to consider, corresponding to the many qualitatively different spacetimes shown in Figs.~\ref{diags1} and \ref{diags2}. In each case, the tunneling probability in the WKB approximation is given by
\begin{equation}\label{probR1toR2}
P \left( R_{1} \rightarrow R_{2} \right) = \left| \frac{\Psi\left( R_{2} \right)}{ \Psi\left( R_{1} \right) } \right|^2 \simeq e^{2i\Sigma_{0}\left[R_{2}-R_{1}\right]}, 
\end{equation}
where $\left[R_{2} - R_{1}\right]$ represents evaluation between the two turning points of the classical motion, and $\Sigma_{0}$ is obtained by integrating Eq.~\ref{dSigma}. The plan of attack is to split the integral into three parts: one over the interior of the bubble, one over the exterior, and one in the neighborhood of the wall.  We thus write:
\begin{equation}
i \Sigma_{0} = F_{I}\left[R_{2} - R_{1} \right] + F_{O}\left[R_{2} - R_{1} \right]  + F_{w} \left[R_{2} - R_{1} \right].
\end{equation}

The integrals $F_{I}$ and $F_{O}$ are found by holding $r_{w}$ and the geometry in the neighborhood of the wall fixed, while allowing nontrivial variation of $L$ and $R$ in the interior and exterior spacetimes. Following FMP, we will integrate $L$ along a path of constant $R$ to the boundary of the classically allowed/forbidden region, and then integrate along this boundary to the desired configuration of $L(r), R(r)$. The momenta vanish along this second leg, and so the integral will be of $\pi_{L}$ over $L$
\begin{eqnarray}
F_{I} &=& \int_{0}^{\hat{r}} dr \int dL (\pm \pi_{L}) \\ 
&=&  \pm \int_{0}^{\hat{r}} dr \left[ i \pi_{L} - RR' \cos^{-1} \left(\frac{R'}{L  a_{\rm ds}^{1/2}}\right) \right]. \nonumber
\end{eqnarray}
Note that there is an ambiguity in the sign. This comes from the fact that the constraints (Eq.~\ref{piLsmall} and \ref{piLbig}) are second order in the momenta, and so we must account for both the positive and negative roots. To keep track of this ambiguity, we will define a variable $\eta \equiv \pm1$ with $\sqrt{\pi_{L}^{2}} = \eta \pi_{L}$. We shall have more to say about this issue later.

At the turning point, $\pi_{L}$ vanishes. The integral evaluated between the two turning points is then
\begin{equation}\label{FI}
F_{I} \left[ R_{2}-R_{1} \right] =  \eta \int_{R_{1}}^{R_{2}} dR R \cos^{-1} \left(\frac{R'}{L  a_{\rm ds}^{1/2}}\right) 
\end{equation}

The integral outside the bubble wall ($r > r_{w}$) is given by
\begin{equation}
F_{O} =  \eta \int_{r_{w}}^{\infty} dr \left[ i \pi_{L} - RR' \cos^{-1} \left(\frac{R'}{L  a_{\rm sds}^{1/2}}\right) \right]
\end{equation}
which evaluated between the two turning point becomes
\begin{equation}\label{FO}
F_{O}\left[R_{2} - R_{1}\right] =  \eta \int_{R_{1}}^{R_{2}} dR R \cos^{-1} \left(\frac{R'}{L  a_{\rm sds}^{1/2}}\right) 
\end{equation}

At the turning point, $R'$ inside and outside of $r_{w}$ is given by solving Eqs. \ref{piLsmall} and \ref{piLbig} for $R'$:
\begin{equation}
R'(r_{w} - \epsilon) = \pm L a_{\rm ds}^{1/2}, \ \ \ \ R'(r_{w} + \epsilon) = \pm L a_{\rm sds}^{1/2}.
\end{equation}
Therefore, the inverse cosine in the integrals of Eq.~\ref{FI} and \ref{FO} are either $0$ when $R'$ is positive or $\pi$ when $R'$ is negative. To perform these integrals, imagine moving the wall along the tunneling hypersurface ($t=0$) between the two turning points (for an example, see Fig.~\ref{tunnel}). The sign of $\beta$ is positive if the coordinate radius $r$ is increasing in a direction normal to the wall and negative if it is decreasing. Therefore, the sign of $R'$ is equal to the sign of $\beta$ as one moves along the tunneling hypersurface, and the integrals Eq.~\ref{FI} and~\ref{FO} will be zero in regions of positive $\beta$ and $\pi$ in regions of negative $\beta$. 

Shown in table \ref{tableworm} are the values of $F_{O}$ and $F_{I}$ for all of the possible transitions where the unbound solution is to the left, on the conformal diagram, of the bound solution (for example, the process shown in Fig.~\ref{tunnel}), which in all cases but $B>3 (A-1)$ with $M>M_S$ (the mass at which $\beta_{sds}$ changes sign on the effective potential) occurs through a wormhole (for $B>3 (A-1)$ with $M>M_S$, the most massive bound and unbound solutions can both be behind a worm hole). We will  refer to these solutions as L(eft) tunneling geometries. These were the solutions studied by FGG and FMP, but we have seen above that there are actually many other allowed processes due to the non-compact properties of the SdS spacetime. These are tunneling processes where the unbound solution lies to the right of the bound solution on the conformal diagram, which we will refer to as R(ight) tunneling geometries. The values of the integrals $F_{I}$ and $F_{O}$ in this case are shown in table \ref{tablenoworm}. In all cases except for $B > 3 (A-1)$ with $M>M_S$, the bubble wall exits the cosmological horizon (whereas the L tunneling geometries went through a wormhole), as in Fig.~\ref{tunnelnoworm} (for $B>3 (A-1)$ with $M>M_S$, the bubble wall traverses a wormhole and cosmological horizon).  

There still is one more integral to evaluate, which allows for the variation of the geometry at the position of the wall 
\begin{widetext}
\begin{eqnarray}\label{Fw}
F_{w} \left[ R_{2} - R_{1} \right] &=& \int_{R_{1}}^{R_{2}} dR_{w} R_{w} \left[ \cos^{-1}\left[\frac{6M + 3 k^2 R_{w}^{3} - R_{w}^{3} \left(\Lambda_{-} - \Lambda_{+}  \right) }{6k R_{w}^{2} a_{\rm ds}  }  \right] \right.  \nonumber \\   && \left. - \cos^{-1}\left[\frac{6M - 3  k^2 R_{w}^{3} - R_{w}^{3} \left(\Lambda_{-} - \Lambda_{+}  \right) }{6k R_{w}^{2} a_{\rm sds}  }  \right]   \right].
\end{eqnarray}
\end{widetext}
We have been unable to find an analytic expression for this integral, and so have evaluated it numerically. 
 
Putting everything together, we can evaluate the tunneling exponent for the various cases shown in tables~\ref{tableworm} and \ref{tablenoworm}. Shown in Fig.~\ref{nosub} is an example of $2 i \Sigma_{0}$ for both the L (blue dashed line) and R (red solid line) tunneling geometries with $3 (A-1) < B < A + 3$ ($A=1$, $B=6$), where we have taken $\eta=+1$. The vertical dashed lines represent the mass scales $M_D$ (left) and $M_{S}$. L tunneling geometries with $M < M_S$ correspond to tunneling through a wormhole. The magnitude of these tunneling exponents is fixed by the inverse bubble wall tension squared ($k^{-2}$), which in geometrical units ranges from $k^{-2} \simeq 10^{102}$ for a tension set by the Weak scale to $k^{-2} \simeq 1$ for a tension set by the Planck scale.

\subsection{High- and low-mass limits}\label{highlow}

\begin{figure}
\begin{center}
\epsfig{file=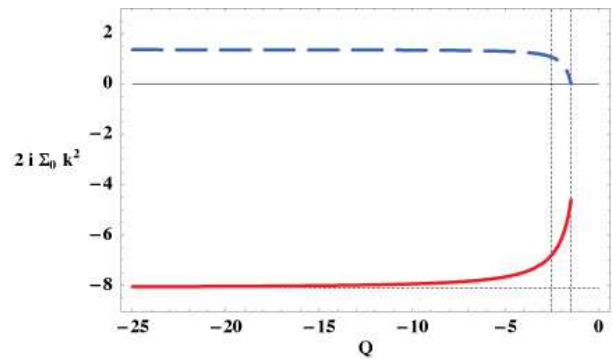,width=8cm}
\caption{Tunneling exponent as a function of $Q$ for ($A=1$, $B=6$) (false vacuum bubbles). The blue dashed line is for the L tunneling geometries, while the red solid line is for the R tunneling geometries. The vertical dotted lines denote the mass scales $M_{D}$ (left) and $M_{S}$ (right) described in Tables~\ref{tableworm} and \ref{tablenoworm}. The horizontal dotted line is at the value of the CDL tunneling exponent (Eq.~\ref{BCDL}). \label{nosub}}
\end{center}
\end{figure}

Note that as the mass increases, the width of the potential barrier that must be crossed decreases (see the potential diagrams in Fig.~\ref{Bgt3Am1}, \ref{Blt3Am1}, \ref{AgtBo3p1}, and \ref{A_6B_5}). We therefore expect that the tunneling exponent (for tunneling through the effective potential) goes to zero at the top of the barrier. However, the tunneling exponent is not always zero at the top of the potential, as can be seen from the tunneling exponent for the R tunneling geometry shown in Fig.~\ref{nosub} (red solid line). To see how this happens, consider a mass slightly below the maximum of the effective potential. The bound solutions are the same for both the L and R tunneling geometries (Solutions 1 or 2), but the unbound solutions to which we are tunneling differ. For a bound Solution 1, we are tunneling to one of the two versions (corresponding to the L or R tunneling geometry) of either Solution 6, 10, or 11 depending on the values of $A$ and $B$ . For a bound Solution 2, we are tunneling to one of the two versions of either Solution 8 or 9. 

In the case where $B > 3(A-1)$ (the situation pictured in Fig.~\ref{nosub}), the most massive L tunneling geometry will have the bound and unbound solutions smoothly merge as the top of the potential barrier is approached. The most massive R tunneling geometry in this case will find the bound and unbound solutions separated by both a black hole and cosmological horizon, and so the tunneling exponent at the top of the potential well will be given by $2 i \Sigma_{0}= \pi \left( R_{S}^{2}-R_{C}^{2} \right)$. This situation is reversed when $B < 3(A-1)$, where the R tunneling geometry will possess the smooth high mass limit, and the most-massive L tunneling geometry will have a non-zero tunneling exponent. 

Now consider the other end of the mass spectrum: the zero mass limit of the two different tunneling geometries. In either case, as the mass is taken to zero, the turning point of the bound solution goes to zero, and the turning point of the unbound solution approaches the nucleation radius of a CDL bubble (see Eq.~\ref{r_0inst}). Even so, there is a fundamental difference between these two solutions when the background spacetime is considered.

As the mass is taken to zero in the L tunneling geometry (corresponding to the FGG mechanism), the worm hole separating the background of the old phase and the bubble of the new phase disappears. This leaves a background spacetime in which absolutely nothing happens, along with a universe containing a CDL bubble which is created from nothing. At least in the zero-mass limit, this means that we are calculating Vilenkin's  tunneling wave function for an inhomogenous universe \cite{Vilenkin:1982de,Vilenkin:1984wp,Vilenkin:1998dn} with the tunneling exponent equal in magnitude to the CDL instanton action (without the background subtraction term). 

This situation is rather strange: if considered one physical system, we have seemingly created new degrees of freedom. It is therefore unclear how we should interpret the tunneling probability; what are we fluctuating out of, and probability per unit what? The massive case seems to create new degrees of freedom as well, since the region to the left of the worm hole (containing large regions of both the old and new phase) in Fig.~\ref{tunnel} does not exist prior to the tunneling event. It is perhaps not so surprising then that Freivogel et. al. \cite{Freivogel:2005qh} have found that when a conformal field theory dual to FGG tunneling from AdS is constructed using the AdS/CFT correspondence, it corresponds to a non-unitary process. 

The zero mass limit of the R tunneling geometry corresponds to the nucleation, in some background, of a CDL true- or false-vacuum bubble. The CDL tunneling exponent (including the background subtraction) can be written as \cite{Garriga:2004nm, Feng:2000if}
\begin{equation}\label{BCDL}
S_{\rm CDL} = \frac{3 \pi}{2} \left[\frac{1}{\Lambda_{+}} \left( 1 - b \alpha_{+} \right) - \frac{1}{\Lambda_{-}} \left( 1 - b \alpha_{-}  \right)   \right],
\end{equation}
where 
\begin{equation}
\alpha_{\pm} = \frac{\Lambda_{+}-\Lambda_{-}}{6k} \mp \frac{k}{2},
\end{equation}
and
\begin{equation}
b = \sqrt{\frac{3}{\Lambda_{-} + 3 \alpha_{-}^{2}}}.
\end{equation}
The horizontal dotted line in Fig.~\ref{nosub} is the value of the CDL tunneling exponent for a particular choice of parameters, and it can be seen that the zero mass limit ($Q \longrightarrow -\infty$) of the R tunneling geometry asymptotes to this. Similar results were found in the case of {\em true}-vacuum bubbles by Ansoldi et. al. \cite{Ansoldi:1997hz}, who were able to reproduce the CDL tunneling exponent using a Hamiltonian formalism. 

It can be seen in Fig.~\ref{nosub}, that the tunneling exponent takes opposite signs for the two tunneling geometries (Fig.~\ref{tunnel} and Fig.~\ref{tunnelnoworm}). For both tunneling probabilities to be less than one, $\eta$ must take opposite signs in each case. We have seen that the zero-mass limit of the L tunneling geometry (FGG mechanism)  corresponds to creation of an inhomogenous universe from nothing. This perspective suggests that the sign choice we are forced to make is a reflection of some quantum-cosmological boundary conditions, since choosing the sign of $\eta$ is tantamount to choosing the growing or decaying wave function in the region under the well. Taking linear combinations of the growing and decaying wave functionals would yield any one of the three existent sign conventions of Hartle and Hawking  \cite{Hartle:1983ai}, Linde \cite{Linde:1983mx}, and Vilenkin \cite{Vilenkin:1982de}. In contrast, the sign choice is rather straightforward for the R tunneling geometries. This process has a clear-cut interpretation in terms of a fluctuation between true- and false-vacuum regions. Thus, we might physically interpret the low CDL probability as the low probability for a downward entropy fluctuation in the background spacetime to occur~\cite{Banks:2002nm}. 

If both tunneling geometries are allowed, we have two processes which correspond to tunneling under the same potential well Eq.~\ref{potential}. It is unclear exactly how one is to interpret this situation, but if it were the case that only one of these two interpretations were valid, there would be a number of important consequences. For example, if the FGG mechanism (L tunneling geometry) is in fact forbidden, then there would be no possible thin-wall false-vacuum bubble nucleation events in Minkowski space. We have also seen above that the bound and unbound solutions will merge into the monotonic solution at the top of the potential for {\em either} the L or R tunneling geometry, but never both. Since in the low mass limit only the R tunneling geometry matches the tunneling exponent for CDL bubbles, if one were to choose between the two mechanisms, either the low or the high mass end of the spectrum would be discontinuous for some range of parameters. We hope to explore these points further in future work.

\begin{center}
\begin{table*}
\caption{$F_{I}\left[ R_{2}-R_{1} \right]$ and $F_{O}\left[ R_{2}-R_{1} \right]$ for the tunneling geometries with the unbound final state lying to the left of the bound initial state (L tunneling geometries). The mass scales indicated can be located on the potential diagrams by identifying $M_{D}$ as the point on the potential where $\beta_{\rm ds}$ changes sign, $M_{S}$ with the point on the potential to the left of the max where $\beta_{\rm sds}$ changes sign, and $M_{\rm SDS}$ with the point on the potential to the right of the max where $\beta_{\rm sds}$ changes sign. \label{tableworm}}
\begin{tabular}{|r|r|r|r|}
\hline
$A$ and $B$&$M$&$F_{I}\left[ R_{2}-R_{1} \right]$&$F_{O}\left[ R_{2}-R_{1} \right]$\\
\hline
$ 3(A-1) < A+3 < B $&$M < M_{D}$&$\frac{\pi}{2}\left(R_{D}^{2} - R_{2}^{2} \right) $&$\frac{\pi}{2}\left(R_{2}^{2} - R_{S}^{2} \right)$\\
\hline
$3(A-1) < A+3 < B $&$M_{D} < M < M_{S}$&$0$&$\frac{\pi}{2}\left(R_{2}^{2} - R_{S}^{2} \right)$ \\
\hline
$3(A-1) < A+3 < B $&$M > M_{S}$&$0$&$\frac{\pi}{2}\left(R_{2}^{2} - R_{1}^{2} \right)$\\
\hline
$3(A-1) < B < A+3  $&$M < M_{S}$&$0$&$\frac{\pi}{2}\left(R_{2}^{2} - R_{S}^{2} \right)$\\
\hline
$3(A-1) < B < A+3  $&$M > M_{S}$&$0$&$\frac{\pi}{2}\left(R_{2}^{2} - R_{1}^{2} \right)$\\
\hline
$A+3 < B < 3(A-1)$ &$M < M_{D}$&$\frac{\pi}{2}\left(R_{D}^{2} - R_{2}^{2} \right) $&$\frac{\pi}{2}\left(R_{2}^{2} - R_{S}^{2} \right)$\\
\hline
$A+3 < B < 3(A-1)$ &$M_{D} < M < M_{SD}$&$0$&$\frac{\pi}{2}\left(R_{2}^{2} - R_{S}^{2} \right)$ \\
\hline
$A+3 < B < 3(A-1)$ &$M<M_{SD}$&$0$&$\frac{\pi}{2}\left(R_{C}^{2} - R_{S}^{2} \right)$\\
\hline
$B < A+3 < 3(A-1)$ &$M<M_{SD}$&$0$&$\frac{\pi}{2}\left(R_{2}^{2} - R_{S}^{2} \right)$\\
\hline
$B < A+3 < 3(A-1)$ &$M>M_{SD}$&$0$&$\frac{\pi}{2}\left(R_{C}^{2} - R_{S}^{2} \right)$\\
\hline
$A > B+3$&$M < M_{\rm CRIT}$&$0$&$\frac{\pi}{2}\left(R_{C}^{2} - R_{S}^{2} \right)$\\
\hline
$A > \frac{B}{3} + 1$&$M < M_{SD}$&$0$&$\frac{\pi}{2}\left(R_{2}^{2} - R_{S}^{2} \right)$\\
\hline
$A > \frac{B}{3} + 1$&$M > M_{SD}$&$0$&$\frac{\pi}{2}\left(R_{2}^{2} - R_{S}^{2} \right)$\\
\hline
$A < \frac{B}{3} + 1$&$M < M_{S}$&$0$&$\frac{\pi}{2}\left(R_{2}^{2} - R_{S}^{2} \right)$\\
\hline
$A < \frac{B}{3} + 1$&$M > M_{S}$&$0$&$\frac{\pi}{2}\left(R_{2}^{2} - R_{1}^{2} \right)$\\ 
\hline
\end{tabular}
\end{table*}
\end{center}

\begin{center}
\begin{table*}
\caption{$F_{I}\left[ R_{2}-R_{1} \right] + F_{O}\left[ R_{2}-R_{1} \right]$ for the tunneling geometries with the unbound final state lying to the {\em right} of the bound initial state (R tunneling geometries). The mass scales indicated can be located on the potential diagrams by identifying $M_{D}$ as the point on the potential where $\beta_{\rm ds}$ changes sign, $M_{S}$ with the point on the potential to the left of the max where $\beta_{\rm sds}$ changes sign, and $M_{\rm SDS}$ with the point on the potential to the right of the max where $\beta_{\rm sds}$ changes sign.  \label{tablenoworm}}
\begin{tabular}{|r|r|r|r|}
\hline
$A$ and $B$&$M$&$F_{I}\left[ R_{2}-R_{1} \right]$&$F_{O}\left[ R_{2}-R_{1} \right]$\\
\hline
$ 3(A-1) < A+3 < B $&$M < M_{D}$&$\frac{\pi}{2}\left(R_{D}^{2} - R_{2}^{2} \right) $&$\frac{\pi}{2}\left(R_{2}^{2} - R_{C}^{2} \right)$\\
\hline
$ 3(A-1) < A+3 < B $&$M_{D} < M < M_{S}$&$0$&$\frac{\pi}{2}\left(R_{2}^{2} - R_{C}^{2} \right)$ \\
\hline
$ 3(A-1) < A+3 < B $&$M > M_{S}$&$0$&$\frac{\pi}{2}\left(R_{2}^{2} - R_{1}^{2} +R_{S}^{2}-R_{C}^{2} \right)$\\
\hline
$3(A-1) < B < A+3  $&$M < M_{S}$&$0$&$\frac{\pi}{2}\left(R_{2}^{2} - R_{C}^{2} \right)$\\
\hline
$3(A-1) < B < A+3  $&$M > M_{S}$&$0$&$\frac{\pi}{2}\left(R_{2}^{2} - R_{1}^{2} +R_{S}^{2}-R_{C}^{2} \right)$\\
\hline
$A+3 < B < 3(A-1)$ &$M < M_{D}$&$\frac{\pi}{2}\left(R_{D}^{2} - R_{2}^{2} \right) $&$\frac{\pi}{2}\left(R_{2}^{2} - R_{C}^{2} \right)$\\
\hline
$A+3 < B < 3(A-1)$ &$M_{D} < M < M_{SD}$&$0$&$\frac{\pi}{2}\left(R_{2}^{2} - R_{C}^{2} \right)$ \\
\hline
$A+3 < B < 3(A-1)$ &$M>M_{SD}$&$0$&$0$\\
\hline
$B < A+3 < 3(A-1)$ &$M<M_{SD}$&$0$&$\frac{\pi}{2}\left(R_{2}^{2} - R_{C}^{2} \right)$\\
\hline
$B < A+3 < 3(A-1)$ &$M>M_{SD}$&$0$&$0$\\
\hline
$A > B+3$&$M < M_{\rm CRIT}$&$0$&$0$\\
\hline
$A > \frac{B}{3} + 1$&$M < M_{SD}$&$0$&$\frac{\pi}{2}\left(R_{2}^{2} - R_{C}^{2} \right)$\\
\hline
$A > \frac{B}{3} + 1$&$M > M_{SD}$&$0$&$0$\\
\hline
$A < \frac{B}{3} + 1$&$M < M_{S}$&$0$&$\frac{\pi}{2}\left(R_{2}^{2} - R_{C}^{2} \right)$\\
\hline
$A < \frac{B}{3} + 1$&$M > M_{S}$&$0$&$\frac{\pi}{2}\left(R_{2}^{2} - R_{1}^{2} + R_{S}^{2} - R_{C}^{2} \right)$\\ 
\hline
\end{tabular}
\end{table*}
\end{center}

Having developed the necessary tools to calculate the exponent for tunneling from bound to unbound vacuum bubbles, we now finish the development of a framework which will allow us to compare the relative likelihood for all thin-walled vacuum transitions to occur. 

\section{Comparison of the Tunneling Exponents}\label{comparison}

Assuming that both the L and R tunneling geometries exist, and that we can choose the overall tunneling exponent to be negative in either case, we now venture to directly compare the tunneling rates for these two processes. In a cosmological setting, we must fluctuate the bound solution which will expand to its turning point and possibly tunnel to one of the unbound solutions. In the absence of a detailed theory of the nature of these fluctuations, we assume that the probability of fluctuating a solution of a given mass is given by the exponential of the entropy change due to the change in the area of the exterior dS horizon in the presence of a mass \cite{Gibbons:1977mu, Albrecht:2004ke}
\begin{equation}
P_{\rm seed} =\exp\left[-\pi \left(\frac{3}{\Lambda_{+}} - R_{C}^{2}  \right)    \right], 
\end{equation}
where $R_{C}$ is the radius of curvature of the cosmological horizon in SdS. 

Once the bound solution has been fluctuated, it must survive until it reaches the turning point of the classical motion. The authors have shown \cite{Aguirre:2005sv} that any solution with a turning point is unstable against non-spherical perturbations. Even quantum fluctuations present on the bubble wall at the time of nucleation will go nonlinear over some range of initial size and mass. Presumably, these asphericities will affect the tunneling mechanism discussed in the previous section, and may be a significant correction to these processes. Seed bubbles can, however, avoid this instability by forming as near-perfect spheres very near the turning point; in the spectrum of possible fluctuations, there will inevitably be some such events.

Assuming that the seed bubble is still reasonably spherically symmetric when it reaches the turning point, the probability to go from empty dS to the spacetime containing an expanding vacuum bubble is given by the product
\begin{equation}\label{P}
P \simeq C P_{\rm seed} e^{2 i \Sigma_{0}} \equiv C e^{-S_{E}},
\end{equation}
where $C$ is a pre-factor that will be neglected in what follows. 

\begin{figure}
\begin{center}
\epsfig{file=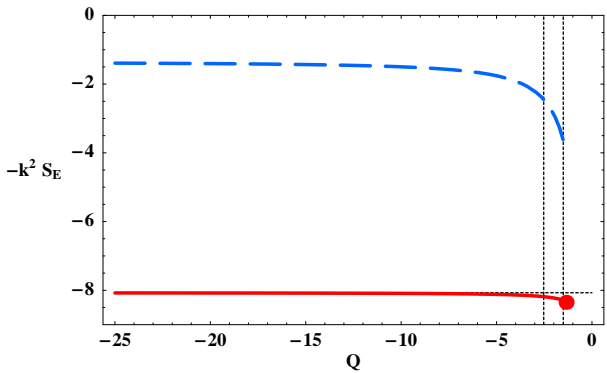,width=8cm}
\caption{The exponent for the creation of a false-vacuum bubble from empty de Sitter as a function of $Q$ for ($A=1$, $B=6$). The blue dashed line is for the L tunneling geometries, while the red solid line is for the R tunneling geometries. The horizontal dotted line is the CDL tunneling exponent. The vertical dotted lines denote the $Q$ corresponding to  $M_{D}$ (left) and $M_{S}$ (right). \label{A1B6sigma}}
\end{center}
\end{figure}

Shown in Fig.~\ref{A1B6sigma} is $-S_{E}$ as a function of $Q$ for ($A=1$, $B=6$), normalized to $k^{-2}$, for both the L tunneling geometries (blue dashed line) and R tunneling geometries (red solid line). In this case, it can be seen that the L tunneling geometries (which pass through the worm hole) are always more probable than R tunneling geometries (which pass through the cosmological horizon). Also, note that the zero mass ($Q \longrightarrow \infty$) solution is in both cases the most probable, even though the width of the potential barrier is largest in this limit.  

We can locate and match the tunneling exponent for thermal activation~\cite{Garriga:2004nm} in Fig.~\ref{A1B6sigma} as the most massive R tunneling geometry (the solution resting on top of the potential in Fig.~\ref{Blt3Am1}), which is denoted by the dot at the far right of the red solid curve. These solutions are bubbles which form in unstable equilibrium between expansion and collapse. We find, in agreement with Garriga and Megevand \cite{Garriga:2004nm}, that thermal activation is always sub-dominant to CDL.   

We have seen above that the R tunneling geometry possesses a smooth high-mass limit only for $B < 3 (A-1)$. The post-tunneling spacetime for this range of parameters is Solution 16 (see Fig.~\ref{thermalons}). However, our picture of the spacetime for $B > 3 (A-1)$ is somewhat different than Solution 17 of Fig.~\ref{thermalons}, which is the post-tunneling spacetime found in Ref.~\cite{Garriga:2004nm}. We find instead that the bubble nucleates outside the cosmological horizon (in the process removing a large section of the background de Sitter) as opposed to behind a worm hole (which leaves the background de Sitter space intact). 

\begin{figure}
\begin{center}
\epsfig{file=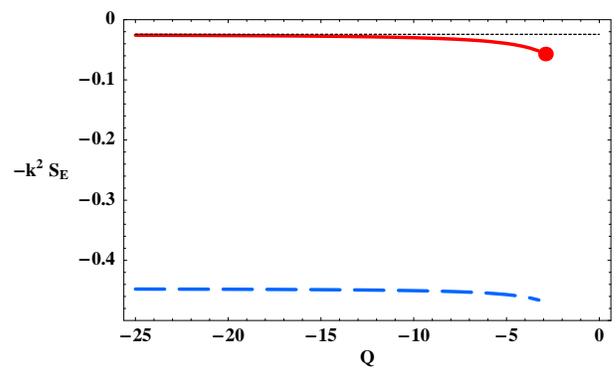,width=8cm}
\caption{Tunneling exponent as a function of $Q$ for ($A=9$, $B=20$) (true-vacuum bubbles). The blue dashed line is for the L tunneling geometries, while the red solid line is for the R tunneling geometries.  \label{A9B20sigma}}
\end{center}
\end{figure}

We have studied examples of the tunneling exponent for all of the possible situations listed in Tables~\ref{tableworm} and \ref{tablenoworm}. The zero mass solution is always the most probable for both the L and R tunneling geometries. Depending on the values of $A$ and $B$, either the L or R tunneling geometries can dominate. Shown in Fig.~\ref{A9B20sigma} is an example of a true-vacuum bubble with ($A=9$, $B=20$); in this case the R tunneling geometries dominate. We can solve for the regions of parameter space where one geometry or another dominates by looking at the zero mass limit. The zero mass limit of the R tunneling geometry is CDL, and the tunneling exponent is given by Eq.~\ref{BCDL} (this includes the background subtraction). The zero mass limit of the L tunneling geometry (FGG) corresponds to the creation from nothing of a universe of the old phase containing a CDL bubble. The tunneling exponent in this case is numerically equal to $3\pi/\Lambda_{+} - S_{CDL}$. Taking the difference of the two tunneling exponents, we find that the L tunneling geometries will be dominant when $2 S_{CDL} > 3\pi/\Lambda_{+}$. 

Depending on the values of the interior and exterior cosmological constant, the picture of vacuum transitions can be very complicated. For comparable cosmological constants, the situation is the most complicated, with both tunneling geometries and all mass scales having tunneling exponents of the same order of magnitude. While one mechanism will dominate, it may not overwhelm the slightly less probable possibilities. In the case where $\Lambda_{+} \ll \Lambda_{-}$, the zero mass limit of the L tunneling geometry (creation of a universe from nothing containing a CDL bubble) dominates. In the case where $\Lambda_{+} \gg \Lambda_{-}$, the zero mass limit of the R tunneling geometry (CDL true-vacuum bubbles) will dominate. 
 
\section{The bottom line}\label{bl}
In the context of the junction condition potentials Figs.~\ref{Bgt3Am1}, \ref{Blt3Am1}, \ref{AgtBo3p1}, and \ref{A_6B_5}, we now have a very organized picture of the types of vacuum transitions which are allowed. At one extreme, corresponding to $Q \rightarrow - \infty$ ($M \rightarrow 0$), we have {\em both} CDL bubble nucleation or the creation of a bubble spacetime from nothing. Moving up the potential in $Q$, we have the L tunneling geometries (FGG mechanism) and/or the R tunneling geometries. These are two-step processes, involving both a thermal fluctuation of the bound solution and a quantum tunneling event through the potential. At the top of the potential, we have the thermal activation mechanism, which is a one step, entirely thermal process. This completes our picture of the possible vacuum transitions, but still leaves unclear which processes actually occur -- as mentioned in the introduction, there has been some question in the community as to whether the FGG process, for instance, is really a valid tunneling mechanism.

The comprehensive semi-classical picture that we have assembled raises a number of important questions in this regard. For instance, we have seen in the derivation of the tunneling exponent that the L and R tunneling geometries require different sign conventions to ensure a well-defined transition amplitude. Since the zero-mass limit of the L tunneling geometry describes the creation of a universe from nothing, this sign choice may well be connected with the notorious sign ambiguity in quantum cosmology~\cite{Vilenkin:1998dn}. However, at the current level of treatment of the problem, there does not seem to be any well defined reason to choose one sign convention over the other, or to allow both. 

There is also the question of how to reconcile the high- and low mass-limits of the L and R tunneling geometries. We have seen that the zero-mass limit of the R tunneling geometry always describes the nucleation of true- or false-vacuum CDL bubbles. It is therefore tempting to use this as evidence that the L tunneling geometries are not allowed. However, in a number of cases the high-mass limit of the R tunneling geometry is discontinuous in the sense that the pre-tunneling bound solution does not approach the post-tunneling unbound solution as the top of the effective potential is reached. In these same cases, the high-mass limit of the L tunneling geometry {\em is} continuous. Thus, even though the low-mass limit of the L tunneling geometry is rather strange (the creation of a universe from nothing), the high-mass limit seems completely reasonable. This complicates any hope of ruling out all L or all R tunneling geometries based on the reasonableness of the high- and low-mass limits of the effective potential.

There is also the problem that the tunneling geometry for some of the processes is not a manifold~\cite{Farhi:1989yr,Fischler:1990pk}, and it is unclear that such metrics should be included in the path integral.  The methods of FGG~\cite{Farhi:1989yr} can be straightforwardly extended to show that the L tunneling geometry is never a manifold. That is, the Euclidean interpolating geometry between the pre- and post-tunneling states always has a degenerate metric. Analyzing the R tunneling geometries is much more involved because both the black hole and cosmological horizons come into play (requiring two coordinate patches), so we will defer a complete analysis of the L and R tunneling geometries to a separate follow-up paper~\cite{Aguirre:xi}.

The unfortunate bottom line, then, is that while the relation between the various nucleation processes is much clearer, the question of which ones actually occur remains  open.

\section{Conclusions}\label{conclusions}

We have catalogued all possible spherically symmetric, thin-wall, one-bubble (true- and false-vacuum) spacetimes with positive cosmological constant and have provided an exhaustive list of the possible quantum transitions between these solutions. Although there are undoubtedly many more possibilities as one relaxes the assumptions of spherical symmetry and a thin wall, this analysis should provide guidance in searching for more realistic processes.

The effective potentials of the junction condition formalism which were used to construct this catalog clearly indicate the existence of a region of classically forbidden radii separating bound solutions from unbound solutions. There are seemingly two processes which correspond to quantum tunneling through this same region, which we refer to as the L and R tunneling geometries. Both processes begin with a bound solution, which might be fluctuated by the background dS spacetime as we have assumed in Section~\ref{comparison}. This bound solution then evolves to its classical turning point, where it has a chance to tunnel to an unbound solution, which is typically either through a wormhole in the case of the L tunneling geometries (the Farhi-Guth-Guven, or FGG, mechanism) or through a cosmological horizon in the case of the R tunneling geometries.

The R tunneling geometries without a wormhole have a relatively clear interpretation in terms of the transition of a background spacetime to a spacetime of a different cosmological constant. Indeed, the zero-mass limit corresponds exactly to the nucleation of true- and false-vacuum CDL (Coleman-De Luccia) bubbles, correctly reproducing the radius of curvature of the bubble at the time of nucleation, as well as the tunneling exponent. 

The L tunneling geometries (FGG mechanism) have a rather perplexing interpretation, which is most clearly seen by studying the zero mass limit. This corresponds to absolutely nothing happening in the background spacetime, while a completely topologically disconnected universe containing a CDL bubble of the new phase is created from nothing. The massive L tunneling geometries also have an element of this creation from nothing. Before the tunneling event, there is no wormhole, but after the tunneling event, there is a wormhole behind which is a large (eventually infinite) region of the old phase surrounded by a bubble of the new phase. It is unclear how we are to interpret this as the transition of a background spacetime to a spacetime of a different cosmological constant, since the background spacetime remains completely unaffected save for the presence of a black hole.

We have found that the sign of the Euclidean action is opposite for the L and R tunneling geometries, and while the second order constraints on the momenta introduce a sign ambiguity, it is unclear how to correctly fix the signs in light of the existence of two seemingly different processes for tunneling in the same direction through the same potential. A complete explanation of these processes may well require the resolution of some very deep problems in quantum cosmology.

If we take the stance that the L and R tunneling geometries are in competition as two real descriptions of a transition between spacetimes with different cosmological constants,  then we must directly compare their relative probabilities. We have shown in Section~\ref{comparison} that the zero-mass solution is always the most probable for either the L or R tunneling geometries, and that the L tunneling geometry will be dominant when $2 B_{CDL} > 3\pi/\Lambda_{+}$. Therefore, if one is considering drastic transitions of the cosmological constant, the zero-mass FGG mechanism will be the dominant mechanism for upward fluctuations and the nucleation of true-vacuum CDL bubbles will be the dominant mechanism for downward fluctuations. This situation upsets the picture of fluctuations in the cosmological constant satisfying some kind of detailed balance \cite{Lee:1987qc, Banks:2002nm}. 

It does, however, help to explain how spawning an inflationary universe from a non-inflating region might be a feasible cosmology \cite{Albrecht:2004ke}. In the picture that we have presented, both the L and R tunneling geometries are constructed by carving some volume out of the background spacetime and filling it with the new phase. The size of this region is in some sense a measure of how special the initial conditions for inflation are. In the case of the R tunneling geometries, a huge number of the states of the background spacetime must be put into the false vacuum at high cost in terms of the probability of such a fluctuation occurring~\cite{Banks:2002nm}. The L tunneling geometries avoid this cost by fluctuating new states already in the false vacuum (seemingly a non-unitary process as discussed by Frievogel et. al.~\cite{Freivogel:2005qh}), with the result that beginning inflation is no longer prohibitively difficult. The question of how much of the background spacetime must make the transition to the false vacuum is therefore crucial to determining exactly how special the initial conditions for inflation are. Unfortunately, detailed balance and this resolution of the paradoxes associated with the initial conditions for inflation are seemingly incompatible, but hopefully future work will yield further insight into the old but still interesting theory of vacuum transitions.

\begin{acknowledgments}
The authors wish to thank A. Albrecht, T. Banks, M. Dine, S. Gratton, and A. Shomer for their assistance in the development of this work.
\end{acknowledgments}


\bibliography{tunneling}

\end{document}